\documentclass[journal,final,letterpaper,twoside,onecolumn,12pt]{IEEEtran}

\usepackage[noadjust]{cite}      

\usepackage{graphicx}  

\usepackage{subfigure} 

\usepackage{url}       

\usepackage{amssymb}

\newtheorem{thm}{\bf Theorem}

\newtheorem{cor}{\bf Corollary}

\newtheorem{lemma}{\bf Lemma}

\newtheorem{example}{\bf Example}

\begin{document}

\title{An MSE Based Transfer Chart to Analyze Iterative Decoding Schemes}
\author{Kapil~Bhattad,~\IEEEmembership{Student Member,~IEEE}
and Krishna~R.~Narayanan,~\IEEEmembership{Member,~IEEE}\\
Department of Electrical Engineering, Texas A\&M University,
College Station, USA\\
Email: kbhattad@ee.tamu.edu, krn@ee.tamu.edu}

\maketitle

\begin{abstract}
An alternative to extrinsic information transfer (EXIT) charts
called mean squared error (MSE) charts that use a measure related
to the MSE instead of mutual information is proposed. Using the
relationship between mutual information and minimum mean squared
error (MMSE), a relationship between the rate of any code and the
area under a plot of MSE versus signal to noise ratio (SNR) is
obtained, when the log likelihood ratio's (LLR) can be assumed to
be from a Gaussian channel. Using this result, a theoretical
justification is provided for designing concatenated codes by
matching the EXIT charts of the inner and outer decoders, when the
LLRs are Gaussian which is typically assumed for code design using
EXIT charts. Finally, for the special case of AWGN channel it is
shown that any capacity achieving code has an EXIT curve that is
flat. This extends Ashikhmin {\em et al}'s results for erasure
channels to the Gaussian channel.
\end{abstract}


\section{Introduction}
\label{sec:Introduction}

An Extrinsic Information Transfer (EXIT) chart is an insightful
and extremely useful tool to analyze iterative decoding schemes.
In an EXIT chart, the mutual information transfer characteristics
of the component decoders is plotted to study the convergence
behavior graphically.

\begin{figure} \centering
\includegraphics[]{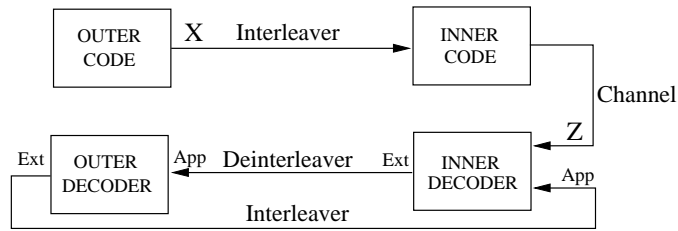}
\caption{Serial Concatenation Scheme}
\label{fig:serial_concatenation}
\end{figure}

Consider a serial concatenation of convolutional codes shown in
Fig. \ref{fig:serial_concatenation}. For this case, EXIT charts
have the following two properties. One, the EXIT curve of the
inner code should lie above the EXIT curve (after reflecting about
the line $y=x$) of the outer code for the iterations to converge
to the correct codeword. Two, the area under the EXIT chart is
related to the rate of the code. If the {\em a priori} information
is assumed to be from an erasure channel, Ashikhmin {\em et al}
\cite{Ashikhmin} showed that for any code of rate $R$, the area
under the exit curve is $1-R$. Based on these properties it is
easy to see that an optimum code can be designed by matching the
EXIT charts. Recently, this technique has been used to design
codes that work well with iterative decoding/signal processing
\cite{matching}. An EXIT chart is usually plotted assuming that
the {\em a priori} LLRs have a Gaussian distribution.  But so far
the area property has been proved only for the erasure case.
Therefore designing codes by matching EXIT charts for Gaussian
{\em a priori} LLRs does not have a theoretical justification,
although it appears to work well in several cases.

In this paper, we define a new measure based on the mean squared
error (MSE) instead of mutual information, and describe an MSE
chart similar to an EXIT chart.  For this new measure, when the
{\em a priori} information is from an AWGN channel, we
theoretically prove an area property that is similar in flavor to
the area property of EXIT charts in erasure channels. We then use
this result to prove that matching of the MSE transfer curves of
the component decoders is optimal when both the {\em a priori} and
extrinsic LLRs are Gaussian. This result is then extended to prove
that EXIT chart matching is also optimal. The proof is based on
the recent result of Guo, Shamai and Verdu \cite{Guo} that relates
the information rate to MMSE and it shows the utility of Guo {\em
et al}'s fundamental result.

We use the area properties derived for the MSE chart to show that
for an AWGN channel, the EXIT chart of a capacity achieving code
is flat. This has recently been proved by Peleg {\em et al} in
\cite{Peleg}. However, the proof in this paper is slightly
different from theirs.

In \cite{Tuchler}, several different measures used to analyze
iterative decoding were studied and it was concluded that some
measures were robust to different channels. However, in order to
compute these measures knowledge of the transmitted bits was
required and, hence, could not be done at the receiver.  We show
that the measure proposed here is robust and can be computed
without knowledge of the transmitted bits.

The paper is organized as follows. In section \ref{sec:Notation}
we present the notation used in this paper. In section
\ref{sec:Measures} we outline some existing measures and propose a
new measure. In section \ref{sec:Area_Property} we show the area
property. We prove the optimality of matching for Gaussian LLRs in
section \ref{sec:Optimality_Matching}. In section
\ref{sec:Memoryless_Channels} we prove that the EXIT chart of
capacity achieving codes is flat. We summarize our results in
section \ref{sec:Conclusion}.

\begin{figure}
\centering
\includegraphics[width = 3.5in]{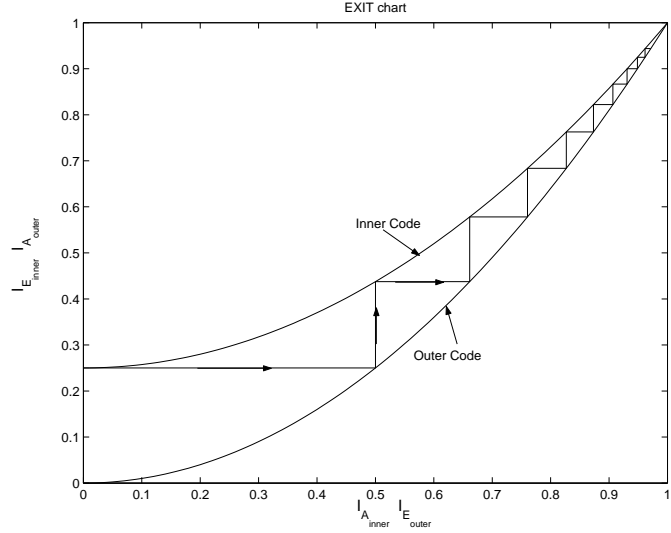}
\caption{Exit Chart} \label{fig:EXIT_chart_example}
\end{figure}

\section{Notation}
\label{sec:Notation} We use $\vec X$ to represent a vector and
$X_1, \ldots, X_n$ to denote its elements. We denote a set
containing elements $X_i, \ldots, X_j$ by $X_i^j$. We use $|{\vec
X}|$ to denote $(\sum X_i^2)^{0.5}$. We use $\phi({\vec X}|{\vec
Y})$ to denote the average minimum mean squared error in
estimating the elements of ${\vec X}$ given ${\vec Y}$, that is
$\phi({\vec X}|{\vec Y}) = \frac{1}{n}E_{{\vec X},{\vec Y}}
\left[\left|{\vec X} - E_{{\vec X}|{\vec Y}}[{\vec X}|{\vec
Y}]\right|^2\right]$. We drop the subscript in the expectation
operator $E[\cdot]$ whenever it is unambiguous.

For the AWGN channel $Y = \sqrt\gamma X + N$, with $X\in\{+1,-1\}$
with $P(X=1) = p$, and $N$ is a Gaussian random variable with zero
mean and unit variance, we use $I_2(\gamma, p)$ to denote the
mutual information between $X$ and $Y$ and $\phi(\gamma, p)$ to
denote the minimum mean squared error in estimating $X$ from $Y$.
When $p = 0.5$ we represent mutual information and MMSE by just
$I_2(\gamma)$ and $\phi(\gamma)$ respectively.

If $Y$ is the output of an AWGN channel with snr $\gamma$ and
input $X$, to highlight that $\phi(X|Y)$ is a function of $\gamma$
we write it as $\phi(X|Y, \gamma)$. We do not encounter cases
where the snr is unknown in this paper.

We will use ${\lambda}$ and ${\rho}$ to represent the edge
perspective degree profile of the variable nodes and the check
nodes in an LDPC code, where $n$ is the total number of edges, $n
\lambda_i$ is the number of edges connected to degree $i$ bit
nodes and $n \rho_i$ is the number of edges connected to degree
$i$ check nodes.

\section{Measures}
\label{sec:Measures} Consider the serial concatenation scheme and
the corresponding iterative decoder shown in Fig.
\ref{fig:serial_concatenation}. Let $L(x_k)$, $L_{ap}(x_k)$ and
$L_{ext}(x_k)$ be the $\log$ likelihood ratio (LLR), {\em a
priori} LLR, and extrinsic LLR on bit $x_k$. Further, let us
assume that the two component decoders produce true {\em a
posteriori} estimates $L(x_k)$ based on $L_{ap}$ and any other
observation from the channel. It has been observed that the pdf of
$L(x_k)$ can be assumed to be Gaussian with mean $mx_k$ and
variance $2m$, denoted by ${\cal N}(mx_k,2m)$. Based on this
assumption, we plot a curve for each of the decoder blocks. We
assume an {\em a priori} LLR $\sim {\cal N}(mx_k,2m)$ and generate
extrinsic LLR for the inner decoder. We extract some parameter
from these LLRs, $F(L)$, and plot $F(L_{ap})$ against
$F(L_{ext})$. For the outer decoder again we do a similar
computation but plot $F(L_{ext})$ against $F(L_{ap})$. This is
illustrated in Fig. \ref{fig:EXIT_chart_example} for $F(L) =
I(X;L)$, in which case such chart is called an EXIT chart. The
path taken by the iterations is also shown in the curve. It is
clear from the chart that the iterations will converge to the
correct codeword if the curves do not cross each other.

We get different charts depending on the parameter that is
extracted. Some of the measures that have been considered
previously are
\begin{description}
\item[M1] Mutual Information measure used in EXIT charts defined
in \cite{TenBrink} is given by
\begin{equation}
F(L) = I(X;L) \label{eqn:IXY}
\end{equation}

\item[M2] Fidelity measure was defined in \cite{Krishna} as
\begin{equation}
F(L) = \theta = E[x_k \cdot  \tanh(L(x_k)/2)] \label{eqn:fidelity}
\end{equation}

\item[M3] In \cite{Tuchler} a measure $\eta$ was defined as
\begin{equation}
F(L) = \eta = E[L^2(x_k)] \label{eqn:eta}
\end{equation}
\end{description}
In \cite{Tuchler}, it was shown that measures M1 and M2 are robust
and predict the performance of iterative decoding well.  Measure
M3 was proposed as a measure that could be computed without
knowing $x_k$'s and, hence, could be used at the receiver.
However, in \cite{Tuchler}, it is shown that this measure is not
robust.

\subsection{Proposed Measure}
\label{sec:Proposed_Measure}

\begin{description}
\item[M4] We propose a new measure $\phi$
\begin{equation}
F(L) = 1 - \phi = E[\tanh^2(L(x_k)/2)] \label{eqn:phi}
\end{equation}
\end{description}

Any APP decoder computes $L_{ext}(x_k)$ from some channel
observations $Y$ and the {\em a prior} information on bits
$x_1^{k-1}$ and $x_{k+1}^n$. When the APP decoder is a true APP
decoder
\begin{equation}
L_{ext}(x_k) = \log \left( \frac{P(X_k=1|Y,
L_{ap}(x_1^{k-1},x_{k+1}^n))}{P(X_k=-1|Y,
L_{ap}(x_1^{k-1},x_{k+1}^n))} \right)
\end{equation}
The MMSE estimate of $x_k$ given $Y$ and $L_{ap}(x_1^{k-1},x_{k+1}^n)$
is given by
\begin{eqnarray}
{\hat x}_k &=&
P(X_k=1|Y,L_{ap}(x_1^{k-1},x_{k+1}^n))-P(X_k=-1|Y,L_{ap}(x_1^{k-1},x_{k+1}^n))\\
&=& \frac{e^{L_{ext}(x_k)}}{1+e^{L_{ext}(x_k)}} -
\frac{1}{1+e^{L_{ext}(x_k)}} = \tanh(L_{ext}(x_k)/2)
\end{eqnarray}
The MMSE is given by
\begin{equation} E[(x_k-\hat{x}_k)^2] =
1 - E[x_k.\hat{x}_k] = 1 - E\left[(\hat{x}_k)^2\right]
\end{equation}
Therefore we have
\begin{equation}
MMSE = \phi = 1-\theta
\end{equation}

From the definition (\ref{eqn:phi}) it can be seen that M4 can be
computed without knowledge of $x_k$. Since M4 is equal to M2 when
the component decoders are true APP decoders it is robust as well.
Let us denote the transfer chart obtained using measure M4 as an
MSE chart.

It is easy to see that when both the {\em a priori} and extrinsic
information are from erasure channels, the MSE chart and the EXIT
charts become identical. Therefore the area properties derived for
the EXIT charts in the erasure case also apply to the MSE chart.
In the next section we derive some area properties for the MSE
chart in the Gaussian case.

\section{Area Property}
\label{sec:Area_Property} In this section we derive some
relationships between the rate and the MSE curve of the inner and
outer code of the serial concatenation scheme shown in Fig.
\ref{fig:serial_concatenation}. The motivation for the
relationships presented here is the following result by Guo {\em
et al} \cite {Guo} that connects MMSE and mutual information.

For a Gaussian channel $Y = \sqrt{snr}X + N;$where $N \sim {\cal
N}(0,1)$, if $\hat{X}$ is the MMSE estimate of $X$ given $Y$ then
\begin{equation}
\frac{d}{dsnr}I(X;Y) = \frac{\log_2e}{2}\ E[(X-\hat{X})^2]
\label{eqn:guoeqn1}
\end{equation}
Using this result when X is binary we get
\begin{equation}
\frac{d}{d\gamma}I_2(\gamma,p) = \frac{1}{\ln4}\phi(\gamma,p)
\label{eqn:guoeqn2}
\end{equation}

\begin{figure}
\centering
\includegraphics[width = 2.5in]{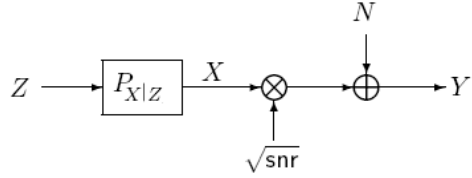}
\caption{General additive noise channel} \label{fig:general_awgn}
\end{figure}



\begin{thm}
Consider a system where ${\vec X}$ is chosen from a code ${\bf C}$
and transmitted over a Gaussian channel with signal to noise ratio
$\gamma$. Let ${\vec Y}$ denote the output of the Gaussian
channel. Let ${\vec Z}$ represent side information available about
${\vec X}$. For this system we have
\begin{eqnarray}
\int_0^{\infty}\phi({\vec X}|{\vec X} \in {\bf C},{\vec Y}, {\vec
Z},\gamma)\ d\gamma = \frac{\ln4}{n} H({\vec X}|{\vec X} \in {\bf
C},{\vec Z})
\end{eqnarray}
where $n$ is the length of the codeword ${\vec X}$.
\label{thm:area_theorem}
\end{thm}
\begin{proof}
This system is similar to the general additive noise channel model
shown in Fig. \ref{fig:general_awgn}. We have
\begin{equation}
I({\vec X};{\vec Y},{\vec Z}|{\vec X} \in {\bf C}) = I({\vec
X};{\vec Z}|{\vec X} \in {\bf C}) + I({\vec X};{\vec Y}|{\vec
Z},{\vec X} \in {\bf C})
\end{equation}
Differentiating both sides with respect to $\gamma$ and noting
that $I({\vec X};{\vec Z}|{\vec X} \in {\bf C})$ is independent of
$\gamma$ we have
\begin{equation}
\frac{d}{d\gamma}I({\vec X};{\vec Y},{\vec Z}|{\vec X} \in {\bf
C}) = \frac{d}{d\gamma} I({\vec X};{\vec Y}|{\vec X} \in {\bf
C},{\vec Z})
\end{equation}
Given ${\vec Z}$ the channel between ${\vec X}$ and ${\vec Y}$ is
Gaussian. By using the relationship derived by Guo {\em et al}
\cite{Guo} we have
\begin{eqnarray}
\nonumber \frac{d}{d\gamma}I({\vec X};{\vec Y},{\vec Z}|{\vec X}
\in {\bf C}) &=& \!\frac{E\!\left[\!\left|{\vec X}-E\!\left[{\vec
X}|{\vec X}\!\in\!{\bf C},{\vec Y},{\vec
Z},\gamma\right]\!\right|^2\!\right]}{\ln4} \\
&=& \frac{n}{\ln4}\phi({\vec X}|{\vec X} \in {\bf C},{\vec Y},
{\vec Z},\gamma)
\end{eqnarray}
Now integrating both sides with respect to $\gamma$ we have
\begin{eqnarray}
 \nonumber \lefteqn{\int_{0}^{\infty} \frac{d}{d\gamma}I({\vec
X};{\vec Y},{\vec Z}|{\vec X} \in {\bf C})\ d\gamma}
\\ \nonumber &&\qquad = I({\vec X};{\vec Y},{\vec Z}|{\vec X} \in {\bf
C})\Big |_{\gamma = \infty} - I({\vec X};{\vec Y},{\vec Z}|{\vec
X} \in {\bf C})\Big |_{\gamma = 0}
\\ \nonumber &&\qquad = H({\vec X}|{\vec X} \in {\bf C}) - I({\vec X};{\vec Z}|{\vec X} \in {\bf C})
\\ \nonumber &&\qquad = H({\vec X}|{\vec Z},{\vec X} \in {\bf C})
\\
\label{eqn:areaproperty1}&&\qquad =
\frac{n}{\ln4}\int_0^{\infty}\phi({\vec X}|{\vec X} \in {\bf
C},{\vec Y}, {\vec Z},\gamma)\ d\gamma
\end{eqnarray}
\end{proof}

Note that in a typical concatenation scheme, ${\vec X}$ is the
input to the inner encoder. However, here we use the term inner
code to refer to a set of constraints satisfied by ${\vec X}$.
This difference will me made clear in example
\ref{ex:outercodeexample}.

\begin{cor}
\label{cor:outercode} For any  code ${\bf C}$ of rate $R$
\begin{equation}
    \int_0^\infty \phi({\vec X}|{\vec X} \in {\bf C},{\vec
Y},\gamma)\ d\gamma = R\ln4
\end{equation}
where ${\vec X}$ represents a length $n$ codeword, ${\vec Y}$
represents the received signal when ${\vec X}$ is transmitted over
an AWGN channel with signal to noise ratio $\gamma$ and
$\phi({\vec X}|{\vec X} \in {\bf C},{\vec Y},\gamma)$ is the MMSE
is estimating ${\vec X}$ given that ${\vec Y}$ is the received
signal when a codeword was transmitted.
\end{cor}
\begin{proof}
Follows from Theorem \ref{thm:area_theorem} when there is no side
information as $H({\vec X}|{\vec X} \in {\bf C}) = R$.
\end{proof}

To plot the transfer characteristic of a component code, it is
assumed that the {\em a priori} information is from a Gaussian
channel. For a true APP decoder, $\tanh(L(X_k)) =
\tanh(L_{ap}(X_k)+L_{ext}(X_k)) = E[X_k|{\vec X}\in {\bf C},{\vec
Y}, \gamma]$. For the outer code in a concatenation scheme, the
$\gamma$ in (\ref{eqn:areaproperty1}), corresponds to the SNR of
the {\em a priori} channel.  Hence, if we plot the MMSE at the
output, $\phi({\vec X}|{\vec X} \in {\bf C},{\vec Y},\gamma) = 1 -
\tanh^2(L(X_k))$, as a function of the {\em a priori} snr then the
area under the curve is equal to the rate of the code times $\ln
4$.
\begin{figure}
\centering
\includegraphics[width = 3.5in]{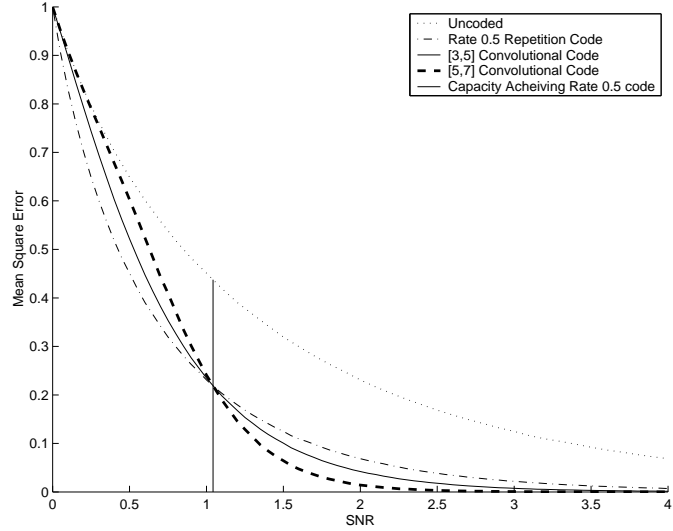}
\caption{MMSE vs SNR } \label{fig:MMSEvsSNR}
\end{figure}

\begin{example}
In Fig. \ref{fig:MMSEvsSNR} we plot the MMSE as a function of SNR
for different rate 1/2 codes. It can be seen that the area under
the MMSE curve for the different codes is nearly the same.
Numerical computations show that the area is nearly $\ln2$.
\end{example}

In context of iterative decoding, corollary \ref{cor:outercode}
provides a nice relationship between the area under the MMSE vs
SNR curve and the rate of an outer code. Theorem
\ref{thm:area_theorem} links the area under the MMSE vs SNR curve
of an inner code to an information theoretic quantity but its
relation to the maximum rate supported is not clear. In the
following lemma, for a special case, when the outer code is chosen
independent of the inner code, we derive a relationship between
the maximum outer code rate supported and the area under the MMSE
vs {\em a priori} snr curve of the inner decoder. Note however,
that this special case is what is typically encountered in
iterative decoding.

\begin{figure}
\centering
\includegraphics[width = 4in]{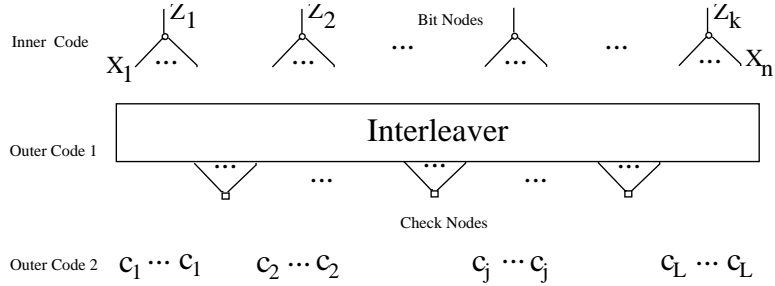}
\caption{Outer Code Example} \label{fig:ldpcexample}
\end{figure}

\begin{example}
\label{ex:outercodeexample} Consider the design of a good LDPC
code designed for an AWGN channel with signal to noise ratio
$\gamma$. We can treat this as a concatenated code where ${\vec
X}$ represent the edges and ${\vec Z}$ the channel observations.
In this case, the inner code represents the restrictions imposed
on ${\vec X}$ by the irregular repeat code (Fig.
\ref{fig:ldpcexample}). The outer code is a single parity check
(SPC) code. We are interested in finding a relationship between
the rate of the SPC outer code and the area under the MMSE chart
for the inner irregular repeat code.  In this case it can be
easily seen that the rate of the outer code $1 - \sum
\frac{\rho_i}{i}$ is bounded above by $ 1-\sum \frac{\lambda_i}{i}
(1-I_2(\gamma)) = 1-\frac 1 n H(X|Z)$. The following Lemma
generalizes this result.
\end{example}

\begin{lemma}
\label{lemma:innercode} If an outer code ${\bf C_{out}}$ is chosen
independent of the inner code ${\bf C_{in}}$, then, the maximum
rate of the outer code that can be used while achieving a
vanishing probability of error is given by $R_{out} \leq 1
 - \frac{1}{n}H({\vec X}|{\vec X} \in {\bf C_{in}},{\vec Z})$ where
${\vec Z}$ represents the channel observation and $n$ represents
length of ${\vec X}$. We will refer to this upper bound as
$R_{outer}^{max}$.
\end{lemma}
\begin{proof}
Let $m$ be the length of the outer codewords and let ${\vec X'}$
represent a length $m$ vector. Consider a sequence $S$ of length
$Nmn$. We say $S \in {\bf C_{in}}$ if $S(ln+1, \cdots, ln+n)$ is a
sequence in ${\bf C_{in}}$ for all $l$. Similarly we say $S \in
{\bf C_{out}}$ if $S(lm+1, \cdots, lm+m)$ is a sequence in ${\bf
C_{out}}$ for all $l$. We say that ${\bf C_{in}}$ and ${\bf
C_{out}}$ are chosen independently if for a random  sequence $S$
the events $S \in {\bf C_{in}}$ and $S \in {\bf C_{out}}$ are
independent, i.e., $P(S \in {\bf C_{in}} \mbox{ and } S \in {\bf
C_{out}})=P(S \in {\bf C_{in}})P(S \in {\bf C_{out}})$.

The number of length $Nmn$ sequences that belong to ${\bf C_{in}}$
is $2^{NmH({\vec X}|{\vec X} \in {\bf C_{in}})}$. Number of length
$Nmn$ sequences that belong to ${\bf C_{out}}$ is $2^{NnH({\vec
X'}|{\vec X'} \in {\bf C_{out}})}$. We have
\begin{equation}
 P(S \in {\bf C_{in}} \mbox{ and } S \in {\bf C_{out}})
 =\frac{2^{NmH({\vec X}|{\vec X} \in {\bf C_{in}})}}{2^{nmN}}\frac{2^{NnH({\vec X'}|{\vec X'} \in {\bf C_{out}})}}{2^{nmN}}
\end{equation}
 and the number of sequences that belong to both ${\bf C_{in}}$ and
 ${\bf C_{out}}$ is
 \begin{equation}
 2^{nmN}\ \frac{2^{NmH({\vec X}|{\vec X} \in {\bf C_{in}})}}{2^{nmN}}\ \frac{2^{NnH({\vec X'}|{\vec X'} \in {\bf C_{out}})}}{2^{nmN}}
 = 2^{NmH({\vec X}|{\vec X} \in {\bf C_{in}})+NnH({\vec X'}|{\vec X'} \in
 {\bf C_{out}})-Nmn}
 \end{equation}

 If with some choice of outer code, the decoder is always able to recover $S$ from the channel
 observations, then the total number of sequences $S$ should be
 less than $2^{NmI({\vec X};{\vec Z}|{\vec X} \in {\bf C_{in}})}$. Therefore we have
\begin{equation}
NmH({\vec X}|{\vec X} \in {\bf C_{in}})+NnH({\vec X'}|{\vec X'}
\in
 {\bf C_{out}})-Nmn \leq NmI({\vec X};{\vec Z}|{\vec X} \in {\bf C_{in}})
 \end{equation}
 which implies
\begin{equation}
\frac{1}{m}H({\vec X'}|{\vec X'} \in
 {\bf C_{out}})\leq 1 - \frac{1}{n}H({\vec X}|{\vec X} \in {\bf C_{in}},Z)
 \end{equation}
\end{proof}

%
%
%
%
%

In the general case (when the outer code is not independent of the
inner code), there seems to be no such relationship.  For example,
consider the LDPC code in Example 2 but consider another outer
code constructed from a good rate $R$ code ($R < I_2(\gamma)$) by
repeating $c_j$, the $j$th coded bit, $d_j$ times, where $d_j$ is
the degree of the $j$th bit node. In this case the rate of the
outer code is $\sum \frac{\lambda_i}{i}R$. Its relationship to
$H(X|Z)$ is not straightforward.

The inner decoder has side information about the coded bits from
the channel output apart from the {\em a priori} information. The
transfer characteristics is obtained by increasing the snr of the
{\em a priori} channel from $0$ to $\infty$. The outer code and
inner code are usually separated by a random interleaver which
makes the inner code and outer code independent. Therefore from
Theorem \ref{thm:area_theorem} and Lemma \ref{lemma:innercode} it
follows that for an inner decoder, the area under the plot of MMSE
at the output against snr of the {\em a priori} channel is equal
to $\ln 4 (1 - R)$, where $R$ is the maximum rate of outer code
supported by the inner code. This can be easily verified for the
following examples.

\begin{example}
Consider an uncoded AWGN channel with signal to noise ratio $SNR$
as an inner code. Let $X$ be the transmitted bit  and let $Z$ be
the received signal. Let $Y$ be the output when $X$ is sent over
another AWGN channel with snr $\gamma$. Clearly, MMSE in
estimating $X$ from $Y$ and $Z$ is same as MMSE in estimating X
from the output of an AWGN channel with an snr of $\gamma + SNR$.
We have
\begin{eqnarray}
\nonumber \int_0^{\infty} \phi(X|Y,Z,\gamma)\ d\gamma &=&
\int_0^{\infty} \phi(\gamma+SNR, p)\ d\gamma \\&=&
\int_{SNR}^{\infty} \phi(\gamma,p)\ d\gamma \\&=& \ln4(H(p) -
I_2(SNR,p)) \label{eqn:example2}
\end{eqnarray}
(\ref{eqn:example2}) follows from (\ref{eqn:guoeqn2}). When $p =
0.5$ we get $\ln4(1 - I_2(SNR))$. $R_{outer}^{max} = I_2(SNR)$ in
this case.
\end{example}

\begin{example}
Consider an uncoded erasure channel with erasure probability
$\epsilon$ as an inner code. Let equiprobable bits $X$ be the
transmitted bits and let $Z$ be the received signal. Let $Y$ be
the output when $X$ is sent over an AWGN channel with snr
$\gamma$. We have $\phi(X|Y,Z,\gamma) = (1-\epsilon)\cdot 0 +
\epsilon \phi(X|Y,\gamma)$. Therefore
\begin{eqnarray*}
 \int_0^{\infty} \phi(X|Y,Z,\gamma)\ d\gamma &=& \epsilon
\int_0^{\infty}
\phi(\gamma)\ d\gamma\\
&=& \epsilon \ln4 \qquad \mbox{ From (\ref{eqn:guoeqn2})}
\end{eqnarray*}
\end{example}
Hence $R_{outer}^{max} = 1-\epsilon$ which is exactly the capacity
of this channel.

\begin{example}
Consider an inner code corresponding to an LDPC code over an AWGN
channel. Let ${\vec X}$ represent the edges, ${\vec Y}$ the {\em a
priori} messages and let ${\vec Z}$ represent the channel
information at the bit nodes. The MMSE for an edge connected to a
bit node of degree $i$ is $\phi(i\gamma + SNR)$. Let
$\{\lambda_i\}$ and $\{\rho_i\}$ represent the degree profile of
the LDPC code in edge perspective. We have
\begin{eqnarray*}
\int_0^{\infty}\!\!\!\!\phi({\vec X}|{\vec X}\!\in\!{\bf C},{\vec
Y},{\vec Z},\gamma)\ d\gamma\!\!\!&=&\!\!\!
\sum_{i=1}^{N_v}\!\lambda_i\! \int_0^{\infty}
\!\!\!\!\phi(i\gamma+SNR)\ d\gamma\\
\!\!\!&=&\!\!\!\ln4 \sum_{i=1}^{N_v}\frac {\lambda_i}{i} (1 -
I_2(SNR))
\end{eqnarray*}
For an LDPC code that works well at $SNR$, we have
\begin{eqnarray*}
1 - \frac{\sum \frac{\rho_i}{i}}{\sum \frac{\lambda_i}{i}} &\leq&
I_2(SNR) \\ \Rightarrow R_{outer} = 1 - \sum \frac{\rho_i}{i}
&\leq&1 - \sum \frac{\lambda_i}{i} (1 - I_2(SNR))
\end{eqnarray*}
\end{example}

It is interesting to compare the area property derived here with
that derived by Ashikhmin {\em et al} in \cite{Ashikhmin}. It was
shown that the area under the EXIT curve, when both the {\em a
priori} and extrinsic information can be modelled to be from
erasure channels, is given by
\begin{equation}
  \mbox{Area} = \left(\frac{1}{n}\sum_{i=1}^n H(X_i| {\vec X} \in {\bf C})\right)^2 \left[ 1 - \frac{H({\vec X}|{\vec Z}, {\vec X} \in {\bf C})}{\sum_{i=1}^n H(X_i| {\vec X} \in {\bf C})}\right]
\label{eqn:ashikhminarea}
\end{equation}
(\ref{eqn:ashikhminarea}) was obtained by modifying equations
(22), (23) in \cite{Ashikhmin} to suit the notation used in this
paper.

In the special case when $H(X_i) = 1$ the area becomes $1 -
\frac{1}{n} H({\vec X}|{\vec Z}, {\vec X} \in {\bf C})$. For an
outer code of rate $R$ the area is therefore is $1-R$. For some
specific inner codes $C_{in}$, it was shown that $1 - \frac{1}{n}
H({\vec X}|{\vec Z}, {\vec X} \in {\bf C_{in}})$ is the maximum
rate of the outer code that can be used in iterative decoding to
achieve error free communication.

In this paper, we have proved that  $1 - \frac{1}{n} H({\vec
X}|{\vec Z}, {\vec X} \in {\bf C_{in}})$ is indeed the maximum
rate of outer code that can be used for reliable communication
when the outer code and inner code are independently chosen. This
makes the area property derived for EXIT charts more concrete.

We note that in the case when $H(X_i|{\vec X} \in {\bf C}) \neq
1$, the simple relationship between the area under the EXIT chart
and rate does not hold. However the relationship between area and
rate of the outer code and the relationship between area and
$R_{outer}^{max}$ for the inner code continue to hold for the MMSE
vs SNR plot.

\subsection{Area Property for MSE chart}

Let us assume that the bits about which information is exchanged
in an iterative decoding scheme (usually the coded bits of the
outer code) are equiprobable. Further, let the {\em a priori} and
the extrinsic information can be modelled as though the bits were
transmitted over an AWGN channel. Let us refer to the SNRs of
these channels as $snr_{ap}$ and $snr_{ext}$. We first note that
if a true APP decoder is employed, MSE is equal to the MMSE. We
denote the MMSE corresponding to the {\em a priori}, the
extrinsic, and the output LLR by $MMSE_{ap}$, $MMSE_{ext}$ and
$MMSE_{out}$ respectively.

We will refer to a plot of  $MSE_{ext}$ versus $MSE_{ap}$ as an
MSE transfer curve. An MSE chart then has two MSE transfer curves,
one for the inner decoder and one for the outer decoder.  The area
properties proved so far are for a plot of the $MMSE_{out}$ (not
$MMSE_{ext}$) versus the $snr_{ap}$. With the Gaussian assumption,
the $MMSE_{out}$ vs $snr_{ap}$ plot can be generated from the MSE
transfer curve using the transformation
$(1-MMSE_{ap},1-MMSE_{ext}) \rightarrow
(\phi(\phi^{-1}(MMSE_{ap})+\phi^{-1}(MMSE_{ext})),\phi^{-1}(MMSE_{ap}))$.
The area properties derived thus apply for the MSE transfer curve
under this transformation. For convenience, we use $\gamma_{ap}$
and $\gamma_{ext}$ to denote $\phi^{-1}(MMSE_{ap})$ and
$\phi^{-1}(MMSE_{ext})$, respectively. We will use subscripts
$inner$ and $outer$ to refer to quantities corresponding to the
inner and outer decoders.

\begin{lemma}
\label{lemma:outercode} If the a {\em a priori} and extrinsic
information can be represented as information from Gaussian
channels then for a rate $R$ code we have $\int_0^{\infty}
\phi(\gamma_{ap} + \gamma_{ext})\ d \gamma_{ext} = (1 - R)\ln4$
\end{lemma}
\begin{proof}
Consider the transfer curve as a continuous curve from $(0,0)$ to
$(1,1)$ by connecting any discontinuity in $\phi(\gamma_{ext})$ by
vertical lines. With every point $(x,y) = (1 - \phi(\gamma_{ap}),1
-\phi(\gamma_{ext}))$ on the transfer curve associate a variable
$z = {x^2+y^2}$. The reason for introducing this variable is to
make it easy to handle the possibility discontinuity of the MSE
transfer curve. It is easy to see that $\gamma_{ap}$ and
$\gamma_{ext}$ are both continuous and increasing functions of $z$
such that $\gamma_{ap}$ and $\gamma_{ext}$ are $0$ at $z = 0$ and
$\infty$ at $z = 2$.
\begin{eqnarray*}
 \ln4 \!\!&=&\!\! \int_{z=0}^{2} \phi(\gamma_{ap} + \gamma_{ext})
\ d
(\gamma_{ap}+\gamma_{ext}) \\
 \!\!&=&\!\! \int_{z=0}^{2} \phi(\gamma_{ap} + \gamma_{ext})
\ d\gamma_{ap}+ \int_{z=0}^{2} \phi(\gamma_{ap} +
\gamma_{ext})\ d\gamma_{ext}\\
 \!\!&=&\!\! \int_{\gamma_{ap}=0}^{\infty}\!\!\!\phi(\gamma_{ap} +
\gamma_{ext})\ d\gamma_{ap}+ \int_{\gamma_{ext}=0}^{\infty}
\!\!\!\phi(\gamma_{ap} +
\gamma_{ext})\ d\gamma_{ext}\\
\!\!&=&\!\! R\ln4+ \int_0^{\infty} \phi(\gamma_{ap} +
\gamma_{ext})\ d\gamma_{ext}
\end{eqnarray*}
\end{proof}

\begin{lemma}
\label{lemma:innercode2} For an inner code when the a {\em a
priori} and extrinsic information can be represented as
information from Gaussian channels then the maximum supported
outer code rate ($R_{outer}^{max}$) is given by $\int_0^{\infty}
\phi(\gamma_{ap} + \gamma_{ext})\ d \gamma_{ext}/(\ln4)$
\end{lemma}
The proof is similar to the proof in the previous lemma.

\begin{example}
Consider a repetition code of rate $1/N$. In this case when the
{\em a priori} information is from an AWGN channel of snr
$\gamma_{ap}$ then the extrinsic information can be modelled as
information from a Gaussian channel of snr $(N-1)\gamma_{ap}$. We
have
\begin{equation}
\label{eqn:uncoded_AWGN}
    \int_{0}^{\infty} \phi(\gamma_{ap}+\gamma_{ext})\ d\gamma_{ext}
    = (N-1)\int_{0}^{\infty} \phi(\gamma_{ap}+(N-1)\gamma_{ap})\ d \gamma_{ap}
    = \frac{N-1}{N}\int_{0}^{\infty} \phi(x)\ dx = \left(1 -
    \frac{1}{N}\right)\ln 4
\end{equation}
This verifies Lemma \ref{lemma:outercode}.
\end{example}

\section{Optimality of Matching}
 \label{sec:Optimality_Matching}

In this section we prove that the MSE curve of the outer code has
to be matched to the MSE curve of the inner code when the
extrinsic information resembles that from an AWGN channel.
\begin{lemma} \label{lemma:formatching} For two codes $C_1$ and
$C_2$ such that $1-MMSE_{ext}^{C_1}(\gamma_{ap}) \leq
1-MMSE_{ext}^{C_2}(\gamma_{ap})\ \forall\ \gamma_{ap}$, $R_1\geq
R_2$ with equality only when the two curves overlap.
\end{lemma}
\begin{proof} We have
\begin{eqnarray*}
MMSE_{ext}^{C_1}(\gamma_{ap}) &\geq& MMSE_{ext}^{C_2}(\gamma_{ap})\\
\Rightarrow \gamma_{ext}^{C_1}(\gamma_{ap}) &\leq& \gamma_{ext}^{C_2}(\gamma_{ap})\\
\Rightarrow \phi(\gamma_{ap}+\gamma_{ext}^{C_1}(\gamma_{ap}))
&\geq& \phi(\gamma_{ap}+\gamma_{ext}^{C_2}(\gamma_{ap}))\\
\Rightarrow R_1&\geq&R_2
\end{eqnarray*}
It is easy to see that equality occurs only when the two curves
overlap.
\end{proof}

From Corollary \ref{cor:outercode} and Lemma
\ref{lemma:innercode2} it follows that a code that is matched
exactly to the channel has a rate equal to the rate supported by
the inner code. Therefore from Lemma \ref{lemma:formatching} it is
easy to see that any outer code whose flipped MSE curve lies below
the inner code and is not matched to the inner code has a rate
lesser than that supported by the inner code.

We note that under the Gaussian assumption the MSE curve and EXIT
curve are related by a one to one function. Therefore since
matching is optimal for MSE chart, it is optimal for EXIT charts.

As a consequence of the results derived so far, under the Gaussian
assumptions, the following properties hold.
\begin{enumerate}
\item With Gaussian assumption on messages from outer decoder to
inner decoder.
\begin{eqnarray}
\frac{1}{\ln4} \int \phi_{inner}^1\ d \gamma_{ap,inner} = 1 - R_{outer}^{max}\\
\frac{1}{\ln4} \int \phi_{outer}^2\ d \gamma_{ext,outer} = 1 -
R_{outer}
\end{eqnarray}
and for the iterative decoder to converge to the correct codeword
$\phi_{inner}^1 < \phi_{outer}^2$. Here $\phi^1$ and $\phi^2$ are
used to denote the MMSE expressed as a function of the {\em a
priori} and the extrinsic snr respectively.

\item With Gaussian assumption on messages from inner decoder to
outer decoder, we have
\begin{eqnarray}
\frac{1}{\ln4}\int \phi_{inner}^2\ d \gamma_{ext,inner} = R_{outer}^{max}\\
\frac{1}{\ln4}\int \phi_{outer}^1\ d \gamma_{ap,outer} = R_{outer}
\end{eqnarray}
and for the iterative decoder to converge to the correct codeword
$\phi_{inner}^2 > \phi_{outer}^1$.
\end{enumerate}

\begin{figure}
\centering
\includegraphics[width = 3.5in]{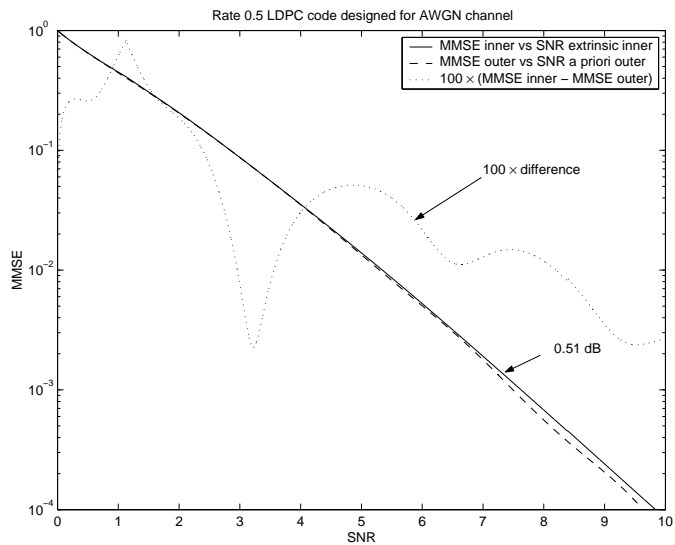}
\caption{Matching Example 1: LDPC code optimized for AWGN channel}
\label{fig:tenbrinkexample}
\end{figure}

\begin{figure}
\centering
\includegraphics[width = 3.5in]{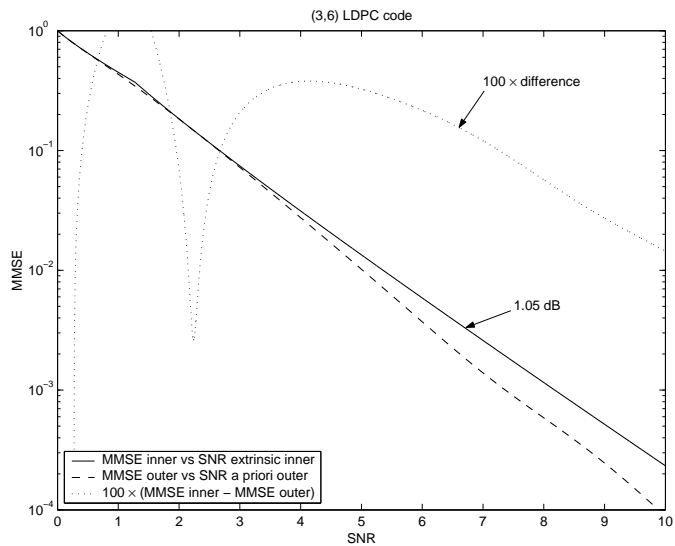}
\caption{Matching Example 2: (3,6) LDPC code} \label{fig:ldpc36}
\end{figure}

Depending on the distribution of the exchanged messages, one of
the above mentioned properties may be used to analyze and design
component codes. For example, in an LDPC code the bit to check
messages closely resembles information from an AWGN channel. In
this case we plot MMSE against SNR extrinsic for the inner code
and against SNR a priori for the outer code. In Fig.
\ref{fig:tenbrinkexample} we plot these curves for a rate 0.5 LDPC
code that was designed using EXIT charts \cite{matching} for an
snr of 0.5dB. The degree profile designed LDPC code is  $\rho_3 =
1$, $\lambda_2 = 0.254$, $\lambda_4= 0.419$, and, $\lambda_{18} =
0.327$. The threshold for a bit error rate of $10^{-4}$ is .55dB.
The threshold predicted using these curves is 0.51dB. In Fig.
\ref{fig:ldpc36} we plot these curve for a (3,6) LDPC code. The
threshold predicted is 1.05dB and the actual threshold is around
1.1dB.

\section{Area Property of Capacity Achieving Codes over AWGN channel}
\label{sec:Memoryless_Channels} The optimality of matching proved
in the previous section assumed that the extrinsic information
resembles information from an AWGN channel. In this section we
prove the optimality of matching for the AWGN channel without
making any assumption on the extrinsic information. We show that
the EXIT curve of any capacity achieving code is flat and is
matched to the channel. It is also seen that the area under the
EXIT curve of any rate $R$ capacity achieving code is equal to $1
- R$.

Consider a capacity achieving binary code ${\bf C}$ of rate $R =
I_2(SNR)$ being transmitted over an AWGN channel with signal to
noise ratio $\gamma$. Since the code decodes perfectly when
$\gamma>SNR$ the MMSE in estimating the transmitted codeword $X$
from the received symbols $Y$ is 0. Therefore from Corollary
\ref{cor:outercode} we get
\begin{eqnarray}
\nonumber
R &=& \frac{1}{\ln4}\int_0^{SNR} \phi({\vec X}|{\vec X} \in {\bf C},{\vec Y}, \gamma)\ d\gamma \\
\nonumber
  &=& \frac{1}{n\ln4}\int_0^{SNR} \sum_{i=1}^{n}\phi(X_i|{\vec X} \in {\bf C},{\vec Y}, \gamma)\ d\gamma \\
\label{eqn:MMSEinequality}
  &\leq& \frac{1}{n\ln4}\int_0^{SNR} \sum_{i=1}^{n}\phi(X_i|Y_i, \gamma)\ d\gamma \\
\nonumber
  &=& \frac{1}{n\ln4}\int_0^{SNR} \sum_{i=1}^{n}\phi(\gamma,P(X_i=1))\ d\gamma  \\
\label{eqn:IMMSEuncodedAWGN}
&=& \frac{1}{n}\sum_{i=1}^{n} I_2(SNR,P(X_i=1)) \\
&\leq& I_2(SNR)
\end{eqnarray}
The inequality in (\ref{eqn:MMSEinequality}) is because MMSE error
in estimating $A$ from both $B$ and $C$ is always less that MMSE
error in estimating $A$ from $B$. It is also easy to prove that
$\phi(A|B,C) = \phi(A|B)$ only when $E[A|B,C] = E[A|B]$.
(\ref{eqn:IMMSEuncodedAWGN}) follows from (\ref{eqn:guoeqn2}).

Since $R = I_2(SNR)$ the inequalities in
(\ref{eqn:MMSEinequality}) and (\ref{eqn:IMMSEuncodedAWGN}) have
to be equalities. Therefore we have $P(X_i=1) = 0.5\ \forall\ i$.
We also have
\begin{equation}
\label{eqn:MMSEequality}
 \int_0^{SNR} \phi(X_i|{\vec X}\in {\bf C}, {\vec Y},
\gamma)\ d\gamma= \int_0^{SNR} \phi(X_i|Y_i, \gamma)\ d\gamma
\end{equation}
Since $\phi(X_i|Y_i, \gamma) \geq \phi(X_i|{\vec X}\in {\bf C},
{\vec Y}, \gamma)$, from (\ref{eqn:MMSEequality}) it follows that
$\phi(X_i|Y_i, \gamma) = \phi(X_i|{\vec X}\in {\bf C}, {\vec Y},
\gamma)$ for almost every $\gamma \in [0,SNR]$.

Now, using the fact that MMSE in both the cases is a decreasing
function of $\gamma$ and the fact that $\phi(X_i|Y_i, \gamma)$ is
continuous, it can be shown that
\begin{equation}
\phi(X_i|{\vec X}\in {\bf C}, {\vec Y}, \gamma) = \phi(X_i|Y_i,
\gamma) \qquad \gamma < SNR, \ \forall\ i
\end{equation}
It is easy to see that the MMSE estimate $E[X_i|{\vec X}\in {\bf
C},{\vec Y}, \gamma] = E[X_i|Y_i, \gamma]$. Therefore for $\gamma
< SNR$ we have
\begin{eqnarray}
\nonumber
\tanh\left(\frac{L_{ap}(X_i)+L_{ext}(X_i)}{2}\right) &=& \tanh\left(\frac{L_{ap}(X_i)}{2}\right) \\
\Rightarrow L_{ext}(X_i) &=& 0
\end{eqnarray}
Therefore $I(X;L_{ext}) = 0$ when $\gamma < SNR$.

When $\gamma > SNR$, $L_{ap}+L_{ext} = +\infty$ when $X = 1$ is
transmitted. Since $L_{ap} <\infty$ we have $L_{ext} = +\infty$
when $X = 1$. Similarly we have $L_{ext} = -\infty$ when $X = -1$.
Therefore $I(X;L_{ext}) = 1$ when $\gamma > SNR$.

We note that in \cite{Peleg} a very similar approach has been used
to arrive at the same result. The proof presented here though is
simpler and avoids some of the steps in \cite{Peleg}.

\section{Conclusion}
\label{sec:Conclusion}

We proposed a new measure based on MSE for analyzing the
convergence behavior of iterative decoding schemes. This measure
is robust and can be computed without the knowledge of the
transmitted bits. Under Gaussian assumptions, we showed a mapping
from the MSE chart such that for any code the area under the map
is equal to the rate. We used this to prove that curve fitting is
optimum in the MSE chart and then extended it to the EXIT chart
case. For the AWGN channel, without making any assumptions on the
distribution of extrinsic LLRs, we showed that capacity achieving
codes have an EXIT chart that is flat and matched to the channel .

\end{document}